# SPECIAL RELATIVITY FROM THE WAVE PICTURE OF THE LIGHT AND OF THE MATTER

Giovanni Zanella

*Dipartimento di Fisica dell'Università di Padova and Istituto Nazionale di Fisica Nucleare, Sezione di Padova, via Marzolo 8, 35131 Padova, Italy*

**Abstract**
*The properties of the light, the Lorentz transformations, and the relation mass-energy, are introduced using the wave picture of the light and of the massive particles.*

## 1. Introduction

As it is known, A. Einstein based its theory of Special Relativity (SR) on two postulates [1]:
1. *The laws of the physics are the same no matter if referred to one or the other of two systems of co-ordinates (Cartesian) in uniform translatory motion.*
2. *Any ray of light moves in the stationary system of co-ordinates with determined velocity (c) whether the ray is emitted by a stationary or by a moving body.*

Hence,    $c$ = *light path / time interval*,
so    *time interval = light path / c*

and putting conventionally $c=1$, the *light path* expresses the *time interval*.
 Einstein derived from these principles the *Lorentz transformation* (LT), with their contra-intuitive results which refuse the idea of a space and of a time absolute.
 In this paper we shall derive LT, and the main results of SR, also on these further hypotheses, implied in SR:

1. *The empty space is isotropic and homogeneous.*
2. *The empty space is not a "stationary" support of physical events.*
3. *The physical events occurring in the same spatial point and in the same instant (coincident events) are inseparable, in time and space, if viewed from any other reference system.*
4. *The closed wave-front of propagation of a radiation, in the empty space, or the closed outline of a rigid body, are closed in respect to any other reference system.*

## 2. The measurement of the time

The concept of time is associated to some variable and measurable phenomenon of reference viewed in its elapsed history.
The most universal time reference is the path of the light (or other electromagnetic wave) which starts from a point source and travels in the empty space along a graduated axis. About the repetitive phenomena they are improperly used as time reference. Indeed:
- they cannot exhibit by themselves their elapsed history, being requested a counting system;
- the frequency of the cycles must be postulated steady.

Einstein itself, to formulate SR, adopted pulses of light to synchronize ideal clocks displaced in different points of the space, tacitly admitting the constancy of the velocity of the light in the empty space
Hence, it is physically more convenient the adoption of *light clocks* (LC), in the place of mechanical clocks. Indeed, LCs can measure the time as a proportional quantity to the path of a ray of light starting in done instant (Fig.1).
It is worth to underline that the equivalence *light path = time interval* furnishes in turn the constancy of *c*, provided that the time interval be identified on the same path of the ray of light and measured from a same reference system, otherwise the constant *c* cannot be assured.

## 3. Limit of velocity of the bodies

An isotropic source of light generates, in the empty space, spherical waves which have necessarily all the same center in the source (Fig.2), if they are viewed by an observer at rest in the source itself. As a consequence, an observer cannot see, from another reference system, superimpositions of these waves, otherwise such superimpositions would appear also locally, because the coincident events are inseparable.

Suppose now, that $S_o$ and S are two identical reference systems having the parallel Cartesian axes $x_o$ and $x$ in uniform translation along their direction (Fig.3). In particular, the system S has velocity $u$ in respect of $S_o$.

Fig.3 shows the behavior, as viewed from $S_o$, of the waves of light generated from a source at rest in the origin of S, supposing the waves starting when the origins of S and $S_o$ are in coincidence. Now, the light-waves are viewed differently from the two systems. Indeed, the observer at rest in the source sees the waves concentric, while the observer at rest in $S_o$ sees the same waves compressed in the direction of $u$, necessarily without superimpositions.

The consequences of this "necessary" behavior are (Fig.3):
1. The velocity of the source (and in general of any body) cannot exceed that of the light, otherwise the waves of the light can appear superimposed.
2. Waves of unknown nature, travelling at velocity greater than the velocity of the light can exist in principle, but the limit of the velocity of the bodies remains that of the light.
3. Hypothetical waves travelling in the empty space at lower velocity of the light does not exist otherwise the limit of the velocity of the bodies would be the latter.

## 4. The Lorentz transformation (LT)

LT is a system of equations which connects the co-ordinates $t_0$, $x_0$, $y_0$, $z_0$, characterizing an event in $S_0$, with the co-ordinates $t$, $x$, $y$, $z$, characterizing the same event in S.

We can deduce LT with reference to a particular class of events. That is looking to Fig.3 and to Fig.4, which is the symmetrical version:
- $x_o$ represents the distance (dark arrow) from the origin of $S_o$ to the wave front, measured from $S_o$, along the positive $x_o$-axis.
- $x$ represents the distance (dark arrow) from the origin of S to the wave front, measured from $S_o$, along the positive $x_o$-axis.
- $t_o$ represents the time of $S_o$ (same dark arrow of $x_o$, if $c=1$) measured from $S_o$, along the positive $x_o$-axis.
- $t$ represents the time of S (same dark arrow of $x$, if $c=1$) measured from $S_o$, along the positive $x_o$-axis.

- The times $t$ and $t_o$ start when from the origins of S and $S_o$ are coincident.

Obviously, Fig.3 and Fig.4, being equivalent, must be corrected to equalize the co-ordinate $x_0$. So, we introduce a scale factor $\gamma$ on the system S, with the intent to determine $x_0$ in function of $\gamma$.
Therefore, looking to Fig.3, we write

$$x = \gamma(x_o - ut_o) ,  \qquad (1)$$

consequently

$$t = \gamma\left(t_o - \frac{ut_o}{c}\right) = \gamma\left(t_o - \frac{ux_o}{c^2}\right) . \qquad (2)$$

The last term of Eq.(2) is obtained putting $t_o = x_o/c$, being $x_o$ the co-ordinate of the spherical front of the wave of light measured from $S_o$. In other words, for any $x_o$, $t_o$ is the time necessary for a pulse of light to travel the distance $x_o$. Besides, the scale factor is used also for the time, being spatial the scale of the times.
So, multiplying both the sides of Eq.(2) with $u$ and summing Eq.(1) with Eq.(2), we have $x + ut = \gamma x_o (1 - u^2/c^2)$ and

$$x_o = \frac{(x+ut)}{\gamma\left(1 - \frac{u^2}{c^2}\right)} . \qquad (3)$$

Now, the events of S measured from $S_0$ remain invariant, also when we consider the system S at rest and the system $S_o$ moving with velocity $-u$ (Fig.4).
So, with reference to Fig.4, Eq.(1) becomes

$$x_o = \gamma(x + ut) , \qquad (4)$$

being $x$ and $t$ measured from $S_0$.
Thus, if $x_o$ of Eq.(4) and $x_o$ of Eq.(3) must be the same, comparing the two equations we obtain the so-called *Lorentz's factor*

$$\gamma = \frac{1}{\sqrt{1 - \frac{u^2}{c^2}}} \quad , \tag{5}$$

which satisfies the requirement $\gamma=1$ when $u=0$, becoming imaginary if $u>c$, no matter the sign of $u$. In other words, the space and the time of S, if viewed from $S_0$, must be amplified by $\gamma$ along the direction of $u$, for reasons of symmetry.

Being $t_o = x_o/c$, we obtain from Eq.(1)

$$x = \gamma(x_o - ut_o) = \gamma x_o (1 - u/c) = \frac{x_o \sqrt{1 - u/c}}{\sqrt{1 + u/c}}, \tag{6}$$

Eq. (6) can be extended also to the times, that is

$$\frac{x}{x_o} = \frac{t}{t_o} = \frac{\sqrt{1 - u/c}}{\sqrt{1 + u/c}} \quad . \tag{7}$$

Therefore, LT for events of S, measured from $S_o$, are

$$x = \gamma(x_o - ut_o), \quad y = y_o, \quad z = z_o, \quad t = \gamma\left(t_o - \frac{ux_o}{c^2}\right), \tag{8}$$

or, looking to Fig.4

$$x_o = \gamma(x + ut), \quad y_o = y, \quad z_o = z, \quad t_o = \gamma\left(t + \frac{ux}{c^2}\right), \tag{9}$$

where $\gamma$ is the *Lorentz's factor*

The co-ordinates $x_o$, $t_o$, and $x$, $t$, of Eq.(8) and Eq.(9), can be extended to any value, having in mind that the times are always positive and that the events whose temporal differences are inferior to the time of connection with a ray of light, do not respect the principle of causality.

It is remarkable that, when the co-ordinates are referred to the front of the light pulse, the ratio $x_o/t_o$ of Eq.(8), and the ratio $x/t$ of Eq.(9) furnishes always the velocity $c$, no matter the value of $u$.

## 5. Relativistic increase of the energy of the light

Expression (7) permits us to decide the change of the frequency of a light wave emitted at the origin of S, when the origins of S and $S_o$ were coincident, if viewed from an observer at rest in $S_o$ and propagating along the positive *x*-axis. Indeed, with reference to Figs.2, 3 and 4, the number of the crests viewed in S in the time *t* (or in the distance *x*) is the same of those viewed in $S_o$ in the time $t_o$ (or in the distance $x_o$), if the source of light were put in the origin of $S_0$.
Therefore, we can write

$$\frac{ct_o}{\lambda_o} = \frac{ct}{\lambda} \quad \text{and} \quad \nu_o t_o = \nu t \quad , \tag{10}$$

being $\lambda$ and $\lambda_o$ the wavelengths measured with a ruler of $S_o$.
Hence

$$\nu_o ct_o = \nu ct \quad \text{and} \quad \nu_o x_o = \nu x \ . \tag{11}$$

being $\nu$ the frequency of the wave emitted from S, but observed from $S_o$, and $\nu_o$ the frequency of the same wave emitted from the origin of $S_o$, but observed locally.
Hence, using Eq.(7) and Eq.(11), where LT is involved, we obtain

$$\nu = \nu_o \frac{x_0}{x} = \nu_0 \frac{\sqrt{1+u/c}}{\sqrt{1-u/c}} = \gamma(1+u/c)\nu_o \quad , \tag{12}$$

Eq.(12) can be rewritten relatively to the energy of a photon, as

$$E = h\nu = \gamma\left(1+u/c\right)h\nu_o = \gamma E_o + \gamma\frac{E_o u}{c} \quad , \tag{13}$$

where *h* is the *Planck's constant*, and $E_0$ represents the energy of the emitted photon when $u=0$, while the sign of *u* changes if the observer in $S_o$ considers the wave of light propagating along the negative *x*-axis of S.
Now, if we look to the energy of the photon as a whole, the energy contributions $\pm\gamma\frac{E_o u}{c}$, which depend in sign on the orientation of propagation of the wave along the *x*-axis, are self-eliminating. So the intrinsic energy of the photon emitted by a source in motion with velocity $u$ will be simply

$$E = \gamma E_o \quad , \tag{14}$$

where $E_0$ is *not the energy at rest of the photon, but the energy of the photon when it is emitted from a source at rest.*

Hence, Eq.(14) tell us that it is impossible for a source of light (and in general for any body) to reach the velocity of the light, otherwise the energy of the emitted photons would become infinite.

In conclusion, if a photon emitted from S is intended as a wave propagating in both the directions of S, when it is viewed from $S_o$, it increments its energy by a coefficient represented from the *Lorentz's factor*.

Obviously we can extend Eq.(14) to the energy of any number of photons.

**5.1.** *Mass-energy relationship*

Expanding in series the *Lorentz factor* we obtain

$$E = E_o + \frac{E_o u^2}{2c^2} + \frac{3}{8}\frac{E_o u^4}{c^4} + \cdots. \tag{15}$$

Neglecting magnitudes of fourth order and higher orders (or when $u<<c$) we may place

$$E = E_o + \frac{E_o u^2}{2c^2} \quad . \tag{16}$$

The term $\frac{E_o u^2}{2c^2}$ of Eq.(16), and those of higher order, represent the energy transmitted from the motion of the source to the photon.

Supposing isolated the system light-source, if the energy of the system does not depend on the proportion light-matter, the relativistic increment of energy, due to the light energy $E_0$, must be compensated by a loss of mass $\Delta m$ of the mass $m$ of the source, such that

$$\frac{\Delta m u^2}{2} = \frac{E_o u^2}{2c^2} \quad . \tag{17}$$

Hence, the quantity $E_0/c^2$ replaces the loss of mass $\Delta m$ in the energy balance of the system. Therefore, $E_0/c^2$ represents the mass associated to the photon confined in the system light-source.

Being Eq. (17) independent on $u$, we can affirm that the emission of a photon of energy $E_0$, from a massive source, corresponds to a loss of mass

$$\Delta m = \frac{E_o}{c^2}, \qquad (18)$$

where $\Delta m$ can become exactly $m$.

**5.2.** *Relativistic photon energy according to the quantum mechanics*

According to the quantum mechanics [3], a single photon emitted from a source in motion is located somewhere in the region of space around the source and has momentum in the direction of the motion of the source.

Looking to the one-dimensional context described by the systems $S_o$ and $S$, the photon is going partly, with the same probability, into each of the two directions of the axes $x_o$ and $x$. The photon is then in a state of motion given by the superimposition of the two states associated with the two components of probability 0.5. When we determine the energy in one of the components the result is the energy of the whole photon or nothing at all. In other words, the photon is forced entirely *to make a sudden jump* [3] into one of the states of motion by an observation. This *sudden* change is due to the *disturbance* on the state of motion of the photon that the observation necessarily makes.

In practice, the intrinsic energy of a photon emitted by a source in motion will correspond to the weighted mean of the two possibilities offered by Eq.(12), that is

$$E = \frac{h}{2}\gamma\left(1+\frac{u}{c}\right)\nu_o + \frac{h}{2}\gamma\left(1-\frac{u}{c}\right)\nu_o = \gamma h \nu_o = \gamma E_o. \qquad (19)$$

## 6. Relativistic increase of the energy of a massive particle

The association of particles with waves is not restricted to the case of the light, because all kinds of particles are associated with waves, according to the known *de Broglie hypothesis* [4]. Now, we can retain, at the same manner of the light, these matter-waves to be locally symmetrical.

Differently on the source of light, the waves generated by a single particle of matter will be necessarily confined in a finite volume with a standing structure, due to the finite energy associated to the particle. Therefore, we can argue, as in the case of the light, that the number of wave crests in $S_o$ be the same of the wave crests in $S_0$, if the particle were put in the origin of $S_0$.

Therefore, in analogy with Eq.(12), indicating $\nu$ and $\nu_o$ the frequencies, in S and $S_0$, of the matter waves which generate the standing structure, we have

$$\nu = \nu_o \frac{\sqrt{1+u/c}}{\sqrt{1-u/c}} \quad , \tag{20}$$

where the sign of $u$ changes if the wave emitted from S is intended as propagating along the negative $x$-axis.

According to *de Broglie hypothesis* [4] the energy of the wave-particle is proportional to its frequency by the *Planck's constant*, that is

$$h\nu = h\nu_o \frac{\sqrt{1+u/c}}{\sqrt{1-u/c}} = h\nu_o \gamma(1+u/c). \tag{21}$$

So, if $E_o$ is the energy of the particle at rest in $S_o$, its energy $E$, when it moves with velocity $u$ in respect of $S_0$, will be as in Eq. (13)

$$E = \gamma E_o + \gamma \frac{E_o u}{c} \quad , \tag{22}$$

where, as for the photon, the term $\gamma \frac{E_o u}{c}$ represents the energy contribution, dependent (in sign) on the direction of propagation of the wave along the $x$-axis of S.

Therefore, if we look (from $S_o$) to the energy of the matter-wave as a whole, the contributions $\pm \gamma \frac{E_o u}{c}$ are self-eliminating and the intrinsic energy of the massive particle will be expressed as in Eqs.(14) and (15), that is

$$E = \gamma E_o = E_o + \frac{E_o u^2}{2c^2} + \frac{3}{8}\frac{E_o u^4}{c^4} + \cdots . \tag{23}$$

The expression $E=\gamma E_0$ is common both to a photon and to a massive particle, but in the case of a photon $E_0$ represents its energy when it is emitted from a source at rest, while for a massive particle $E_0$ represents the own energy at rest.

For $u \ll c$ we can interpret $E_o \dfrac{u^2}{2c^2} = \dfrac{mu^2}{2}$, being $m$ the mass of the particle.
So we obtain

$$E_o = mc^2 . \qquad (24)$$

In other words, we can accept the equivalence mass-energy only for a body at rest, while if the body travels at velocity $u$ its energy increases, due to kinetic contributions, that is

$$E = \gamma\, E_o = mc^2 + \frac{mu^2}{2} + \frac{3}{8}\frac{mu^4}{c^2} + \cdots = \gamma mc^2 , \qquad (25)$$

where the *Lorentz factor* does not reveal an relativistic increase of the mass, but the kinetic contribution to the energy of the particle.
The relation $E=\gamma mc^2$ confirms us that a massive particle cannot travel at the velocity of the light, otherwise its energy would be infinite.
Besides, in analogy with the photon, we can arrive to Eq. (23) according to quantum mechanics.
Obviously we can extend Eq.(25) to the energy of any number of particles.

## 8. Conclusions

The wave picture of the light and of the matter permits us to reach, in elementary and intuitive manner, the main results of the Special Relativity. Indeed, concepts as space-time, limit of the velocity of the light, Lorentz transformations and relation mass-energy appear as logical necessity.
It also demonstrated that the quantum picture of the particles does not disagree with the wave picture introduced by L. de Broglie [4].
In conclusion, if L. de Broglie utilized the equivalence mass-energy to hypothesize the wave aspect of the matter, in our case we assume the wave aspect of the matter to derive the equivalence mass-energy, associating the relativistic increase of the frequency of the light-waves and of the matter-waves to the increase of the energy.

**Figure captions**

Fig.1. Schematic drawing of a light clock (LC). Two ideal mirrors are displaced orthogonally to a Cartesian *x*-axis to reflect a light ray out coming at *x*=0 and *t*=0. The time *t* (in arbitrary units) is that spent by the light to cover the distance between the mirrors per the number of reflections.

Fig.2. Two-dimensional representation of waves of light emitted from a source in the empty space and viewed from the source itself.

Fig.3. Two-dimensional representation of waves of light emitted from a source in motion the empty space. The source moves with velocity *u* in respect of the system $S_o$ and the waves are viewed from $S_o$.

Fig.4. Two-dimensional representation of waves of light emitted from a source in motion the empty space. The system $S_0$ moves with velocity -*u* in respect of the source S and the waves are viewed from $S_0$.

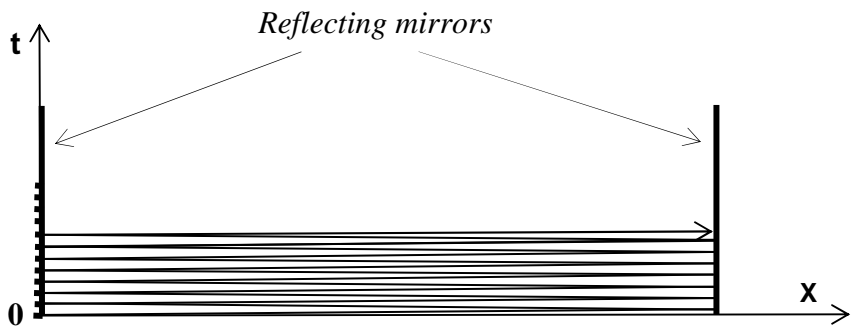

**FIG.1**

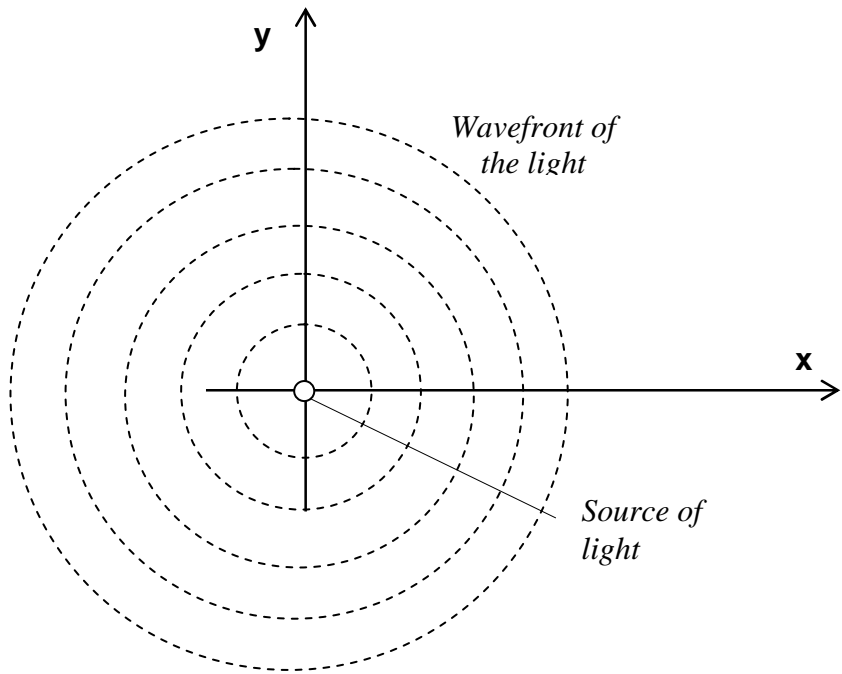

**FIG.2**

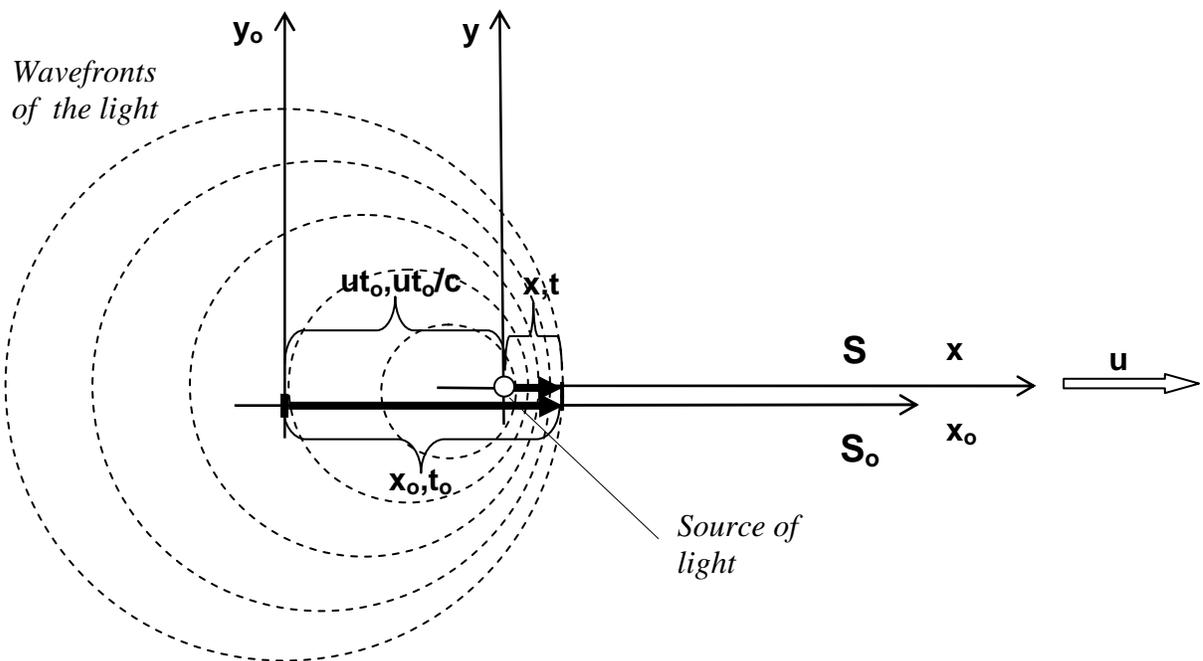

**FIG. 3**

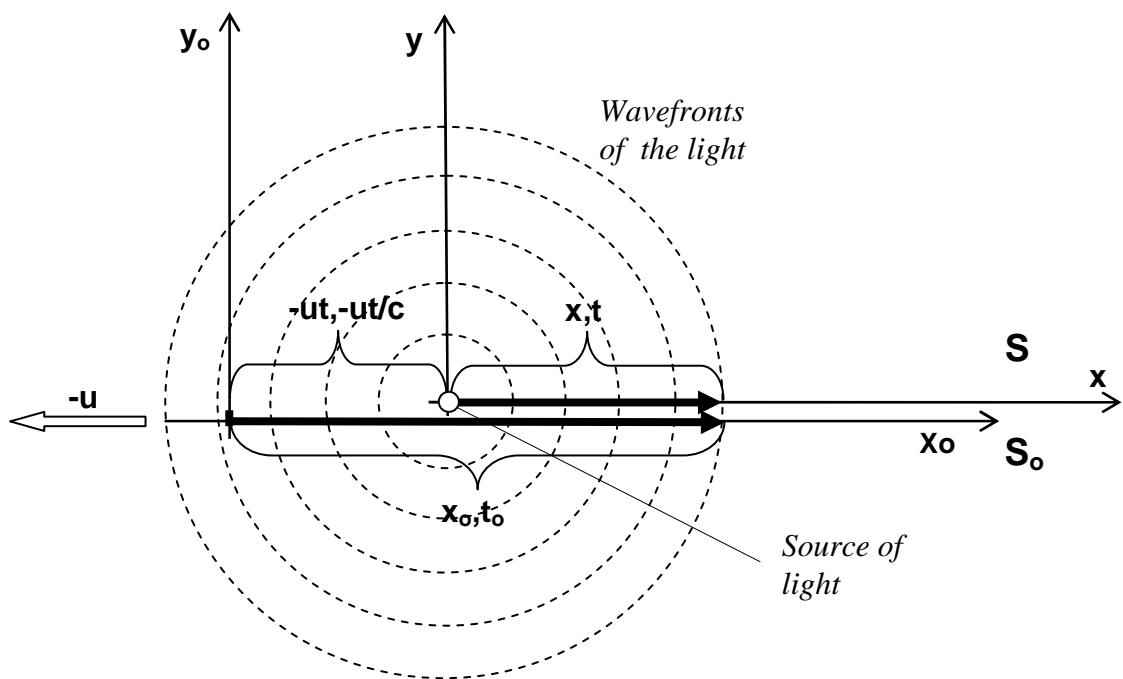

**FIG. 4**